\documentclass[11pt]{article}

\usepackage{amsmath}
\usepackage{amssymb}
\usepackage{extarrows}
\usepackage{graphicx}
 \usepackage{indentfirst}
\usepackage{hyperref}

\renewcommand{\baselinestretch}{1.3}
  \renewcommand{\arraystretch}{1.2}
\begin{document}

 \title{A Note on ``Efficient Algorithms for Secure \\ Outsourcing of Bilinear Pairings"}

  \author{Lihua Liu$^1$ \qquad  Zhengjun Cao$^2$}
  
   \footnotetext{
     $^1$Department of Mathematics, Shanghai Maritime University,
  China.    \textsf{liulh@shmtu.edu.cn}\\ 
   $^2$Department of Mathematics, Shanghai University, Shanghai,
  China.   
    }

 \date{}\maketitle

\begin{quotation}
 \textbf{Abstract}.  We show that the verifying equations in the scheme [Theoretical Computer Science, 562 (2015), 112-121] cannot filter out some malformed values returned by the  malicious servers.  We also remark that the two untrusted programs model adopted in the scheme is somewhat artificial, and discuss some reasonable scenarios for outsourcing computations.

  \textbf{Keywords.} bilinear pairing, outsourcing computation, semi-honest server.
 \end{quotation}

 \section{Introduction}
 Very recently, Chen et al.~\cite{C15} have put forth a scheme for outsourcing computations of bilinear pairings in two
untrusted programs model which was introduced by Hohenberger and Lysyanskaya \cite{HL05}. In the scheme, a user $T$ can indirectly compute the pairing $e(A, B)$ by outsourcing  some expensive work to two untrusted servers  $U_1$ and $U_2$ such that  $A$,  $B$ and $e(A, B)$ are kept secret. Using the returned values from $U_1$, $U_2$ and  some previously stored values, the user $T$ can recover $e(A, B)$.

The Chen et al.'s scheme is derived from the Chevallier-Mames et al.'s scheme \cite{CC10} by storing some values in a table  in order to save some expensive operations such as point multiplications and exponentiations. Besides, the new scheme introduces two servers $U_1$ and $U_2$ rather than the unique server $U$ in the Chevallier-Mames et al.'s scheme. The authors \cite{C15} claim that the scheme achieves the security  \emph{as long as one of the two servers is honest}.
In other word, a malicious server cannot obtain either $A$ or $B$.  Unfortunately, the assumption cannot ensure that the scheme works well,
 because a malicious server  can return some random values  while the user $T$ cannot detect the malicious behavior. As a result, $T$ outputs a false value.

 In this note,  we show that the verifying equations in the scheme \cite{C15} cannot filter out some malformed values returned by the  malicious servers. To fix this drawback, we should specify that the servers are semi-honest.   We also point out that the two untrusted programs model adopted in the scheme is somewhat artificial and  discuss some reasonable scenarios for outsourcing computations.

\section{Review of the scheme}

Let $\mathbb{G}_1$ and $\mathbb{G}_2$ be two cyclic additive groups with a large prime order $q$. Let $G_3$ be a cyclic multiplicative group of the same order $q$. A bilinear pairing is a map $e : \mathbb{G}_1\times \mathbb{G}_2 \rightarrow \mathbb{G}_3$ with the following properties.
(1) Bilinear: $e(aR, bQ) =e(R, Q)^{ab}$ for all $R \in \mathbb{G}_1$, $Q \in \mathbb{G}_2$, and $a, b \in \mathbb{Z}_q^*$.
(2) Non-degenerate: There exist $R \in \mathbb{G}_1$ and $Q\in \mathbb{G}_2$ such that $e(R, Q) \not = 1$.
(3) Computable: There is an efficient algorithm to compute $e(R, Q)$ for all $R \in \mathbb{G}_1, Q\in \mathbb{G}_2$.

 The Chen et al.'s scheme \cite{C15} uses two untrusted servers $U_1, U_2$.  The outsourcer $T$ queries some pairings to the two servers. The scheme can be described as follows.

 \emph{Setup}. A trusted server computes a table \emph{Rand} which consists of the elements of random and independent six-tuple $(W_1, W_2, w_1W_1, w_2W_1, w_2W_2, e(w_1W_1, w_2W_2))$, where $w_1, w_2 \in_R \mathbb{Z}_q^*, W_1\in_R \mathbb{G}_1$, and $W_2\in_R \mathbb{G}_2$.  The table is then loaded into the memory of $T$.

\begin{itemize}
\item{} \emph{Look-up table}. Given $A \in \mathbb{G}_1, B \in \mathbb{G}_2$, where  $A$ and $B$ may be secret or protected and $e(A, B)$ is always secret or protected.  $T$ looks up \emph{Rand} to create
$$ (V_1, V_2, v_1V_1, v_2V_1, v_2V_2, e(v_1V_1, v_2V_2)),$$
 $$(X_1, X_2, x_1 X_1, x_2 X_1, x_2 X_2, e(x_1 X_1, x_2 X_2)),$$
 $$(Y_1, Y_2, y_1Y_1, y_2Y_1, y_2Y_2, e(y_1Y_1, y_2Y_2)).$$

\item{} \emph{Interaction with $U_1$}.  $T$ sends
$\{(A + v_1V_1, B + v_2V_2),   (v_1V_1 + v_2V_1, V_2),  (x_1X_1, x_2X_2),   (y_1Y_1, y_2Y_2)\}$
to  $U_1$. $U_1$  returns
$$\alpha_1=e(A + v_1V_1, B + v_2V_2),\  \delta=e(V_1, V_2)^{v_1+v_2},\  \beta_1=e(x_1X_1, x_2X_2), \ \beta_2=e(y_1Y_1, y_2Y_2).$$

\item{} \emph{Interaction with $U_2$}.  $T$ sends
$\{(A + V_1, v_2V_2),   (v_1V_1, B + V_2),   (x_1X_1, x_2X_2),  (y_1Y_1, y_2Y_2)\}$ to  $U_2$. $U_2$ returns
$$\alpha_2=e(A + V_1, v_2V_2),\ \alpha_3= e(v_1V_1, B + V_2),\  \widehat{\beta_1}=e(x_1X_1, x_2X_2), \ \widehat{\beta_2}= e(y_1Y_1, y_2Y_2).$$

  \item{} \emph{Verification}. $T$ checks that both $U_1$ and $U_2$ produce the correct outputs by verifying that
  $$\beta_1=\widehat{\beta_1}\ \mbox{and}\ \beta_2=\widehat{\beta_2}.$$ If not, $T$ outputs ``error".

    \item{} \emph{Computation}.  $T$  computes  $e(A, B) =\alpha_1\alpha^{-1}_2 \alpha^{-1}_3 \delta \cdot e(v_1V_1, v_2V_2)^{-1}.$
\end{itemize}

\textbf{Remark 1}.  In the original description of the scheme,  the step 2 (see section 4.2 in Ref.\cite{C15}) has not specified any actions. It only explains that the pairing $e(A, B)$  can be composed by the related values. The authors have confused the explanation with steps of the scheme (it is common that a step of a scheme should specify some actions performed by a participator), which makes the original description somewhat obscure.

\section{The checking mechanism in the scheme  fails}

 In the scheme,  to check whether the returned values  $\alpha_1,\,  \delta,\,  \beta_1, \, \beta_2$ and $\alpha_2,\, \alpha_3,\,  \widehat{\beta_1}, \, \widehat{\beta_2}$ are properly formed, the user $T$  has to check the verifying equations
$$\beta_1=\widehat{\beta_1}\ \mbox{and}\ \beta_2=\widehat{\beta_2}.$$
We now want to stress that the checking mechanism \emph{cannot filter out some malformed values}.
 The drawback is due to that the protected values $A, B$ are not involved in the equations at all.

For example, upon receiving $\{(A + v_1V_1, B + v_2V_2),\  (v_1V_1 + v_2V_1, V_2),\  (x_1X_1, x_2X_2), \ (y_1Y_1, y_2Y_2)\}$, $U_1$ picks a random $ \rho\in \mathbb{Z}_q^*$ and returns
$$\alpha_1=e(A + v_1V_1, B + v_2V_2),\  \rho,\  \beta_1=e(x_1X_1, x_2X_2), \ \beta_2=e(y_1Y_1, y_2Y_2)$$
  to $T$. In such case,  we have
  $ \alpha_1\alpha^{-1}_2 \alpha^{-1}_3 e(v_1V_1, v_2V_2))^{-1} \rho =e(A, B)e(V_1, V_2)^{-v_1-v_2}\rho.$
  That means $T$ obtains $e(A, B)e(V_1, V_2)^{-v_1-v_2}\rho$ instead of  $e(A, B)$.

  To fix the above drawback, we have to specify that \emph{both two servers are semi-honest}. The term of semi-honest here means that a server  can copy the involved values and always returns the correct outputs, but cannot conspire with the other server.
  Under the reasonable assumption, the original scheme can be greatly simplified. We now present a revised version of it as follows.
  \begin{itemize}
\item{} \emph{Look-up table}. Given $A \in \mathbb{G}_1, B \in \mathbb{G}_2$, where $A$ and $B$ may be secret or protected and $e(A, B)$ is always secret or protected.  $T$ looks up \emph{Rand} to create
$$ (V_1, V_2, v_1V_1, v_2V_1, v_2V_2, e(v_1V_1, v_2V_2)).$$

\item{} \emph{Interaction with $U_1$}.  $T$ sends
$\{(A + v_1V_1, B + v_2V_2),\  (v_1V_1 + v_2V_1, V_2)\}$
to  $U_1$. $U_1$  returns
$$\alpha_1=e(A + v_1V_1, B + v_2V_2),\  \delta=e(V_1, V_2)^{v_1+v_2}.$$

\item{} \emph{Interaction with $U_2$}.  $T$ sends
$\{(A + V_1, v_2V_2),\  (v_1V_1, B + V_2)\}$ to  $U_2$. $U_2$ returns
$$\alpha_2=e(A + V_1, v_2V_2),\ \alpha_3= e(v_1V_1, B + V_2).$$

  \item{} \emph{Computation}.  $T$ computes
   $e(A, B) =\alpha_1\alpha^{-1}_2 \alpha^{-1}_3 e(v_1V_1, v_2V_2)^{-1} \delta.$
\end{itemize}

\section{The checking mechanism in the Chevallier-Mames et al.'s scheme works well}

As we mentioned before, the Chen et al.'s scheme \cite{C15} is derived from the Chevallier-Mames et al.'s scheme \cite{CC10}.
But the new scheme misses the feature of the checking mechanism in the Chevallier-Mames et al.'s scheme.
We think it is helpful for the later practitioners to explain the feature.

In the Chevallier-Mames et al.'s scheme, the outsourcer $T$ wants to compute the pairing $e(A, B)$ with the help of the untrusted server $U$ such that
$A, B$ and $e(A, B)$ are kept secret. The scheme can be described as follows (See Table 1).

\begin{center}
Table 1: The Chevallier-Mames et al.'s scheme\vspace*{3mm}

\begin{tabular}{|l|c|l|}
  \hline

  The outsourcer $T$  &   & The  server $U$ \\
   \{$P_1 \in \mathbb{G}_1, P_2 \in \mathbb{G}_2$, $e(P_1, P_2)$\}& & \\  \hline
  Input: $A\in \mathbb{G}_1, B\in \mathbb{G}_2$   & & \\
   Pick $g_1, g_2, a_1, r_1, a_2, r_2\in Z_q^*$,  & & \\
   compute $ A + g_1P_1,  B + g_2P_2$ & & Compute \\
   $a_1 A +r_1P_1, a_2B +r_2P_2$  & $\stackrel{(A + g_1P_1, P_2)}{-------\rightarrow}$  & $\alpha_1=e(A + g_1P_1, P_2)$ \\
     and query them.  &$\stackrel{(P_1, B + g_2P_2)}{-------\rightarrow}$   &  $\alpha_2=e(P_1, B + g_2P_2)$   \\
    &$\stackrel{(A + g_1P_1, B + g_2P_2)}{-------\rightarrow}$  &   $\alpha_3=e(A + g_1P_1, B + g_2P_2)$  \\
   Compute & $\stackrel{(a_1 A +r_1P_1, a_2B +r_2P_2)}{-------\rightarrow}$  & $\alpha_4=e(a_1 A +r_1P_1, a_2B +r_2P_2)$  \\
     $ e(A, B) =\alpha_1^{-g_2}\alpha_2^{-g_1}\alpha_3 e(P_1, P_2)^{g_1g_2}$   &  $\stackrel{\alpha_1, \alpha_2, \alpha_3, \alpha_4 }{\leftarrow -------}$  & and return them.  \\
 Check that &&\\
  $ \alpha_4\stackrel{?}{=} e(A, B)^{a_1a_2} \alpha_1^{a_1r_2}\alpha_2^{a_2r_1}$ & & \\
 \qquad  $\cdot e(P_1, P_2)^{r_1r_2-a_1g_1r_2-a_2g_2r_1}$ &   &   \\
 If true, output $e(A, B)$. & & \\
  \hline
\end{tabular}\end{center}

Notice that the true verifying equation is
\begin{eqnarray*}
\alpha_4&=& \left(\alpha_1^{-g_2}\alpha_2^{-g_1}\alpha_3 e(P_1, P_2)^{g_1g_2}\right)^{a_1a_2} \alpha_1^{a_1r_2}\alpha_2^{a_2r_1}
   e(P_1, P_2)^{r_1r_2-a_1g_1r_2-a_2g_2r_1}\\
   &=& \alpha_1^{-g_2a_1a_2+a_1r_2}\alpha_2^{-g_1a_1a_2+a_2r_1}\alpha_3^{a_1a_2} e(P_1, P_2)^{g_1g_2a_1a_2+r_1r_2-a_1g_1r_2-a_2g_2r_1}
   \end{eqnarray*}
   where $\alpha_1, \alpha_2, \alpha_3, \alpha_4$ are generated by the server $U$, and the session keys $g_1, g_2, a_1, r_1, a_2, r_2$ are randomly picked by the outsourcer $T$.

   Clearly, the server $U$ cannot generate the four-tuple $(\alpha_1, \alpha_2, \alpha_3, \alpha_4)$ satisfying the above verifying equation because the exponents
   $$a_1r_2-g_2a_1a_2, \ a_2r_1-g_1a_1a_2,\ a_1a_2,\ g_1g_2a_1a_2+r_1r_2-a_1g_1r_2-a_2g_2r_1$$
   are not known to the server. The intractability of  the above equation can be reduced to the following general challenge:
   $$\mbox{Without knowing a secret exponent}\, \theta,  \mbox{find}\, X, Y \in Z_q^*, \, X\neq 1, Y\neq 1,  \mbox{such that}\, X^{\theta}=Y. $$

\section{The remote and shared servers}

The authors stress that  the two servers $U_1$ and $U_2$, in the real-world applications,  can be viewed as two copies of one advertised software from two different vendors. We would like to remark that the two copies are neither nearby nor private. They must be remote and shared by many outsourcers. Otherwise, the user $T$ equipped with two private copies of one software can be wholly viewed as an \emph{augmented user}.
But the situation is rarely considered in practice.

We now consider the situation that the outsourcer $T$ has to communicate with two remote and shared servers.
If the data transmitted over channels are not encrypted, then an adversary can obtain $A + v_1V_1, B + v_2V_2$ by tapping the communication  between $T$ and $U_1$, and get $v_1V_1,  v_2V_2$ by tapping the communication  between $T$ and $U_2$. Hence, he can recover  $A$ and $B$. Thus, it is reasonable to assume that all data transmitted over channels are encrypted.
 From the practical point of view,  the communication costs (including that of authentication of the exchanged data, the underlying encryption/decryption, etc.) could be far more than the computational gain in the scheme.  The authors have neglected the comparisons between the computational gain and the incurred communication costs.
Taking into account this drawback, we think the scheme is somewhat artificial.

\subsection{A nearby and trusted server}

Girault and Lefranc~\cite{GL05} have described some situations
 in which a chip has only a small computation capability  is connected to a powerful device.
 \begin{itemize}
\item{} In a GSM mobile telephone, the more sensitive cryptographic operations are
performed in the so-called SIM (Subscriber Identification Module), which is already
aided by the handset chip, mainly to decipher the over-the-air enciphered
conversation.

\item{} In a payment transaction, a so-called SAM (Secure Access Module)
is embedded in a terminal already containing a more powerful chip.

\item{} A smart card is plugged into a personal computer,
seeing that many PCs will be equipped with smart card readers in a near
future.
\end{itemize}
We find that in all these situations (a SIM vs. a handset, a SAM vs. a powerful terminal, a smart card vs. a personal computer) the servers are nearby and trusted, not remote and untrusted.

\section{Conclusion}

 The true goal of outsourcing computation in the Chen et al.'s  scheme is to compute bilinear pairings. In view of that pairings spread everywhere in pairing-based cryptograph,  we do not think that the trick of equipping a low capability chip with two untrusted softwares is useful. In practice, we think,  it is better to consider the scenario where a portable chip has access to a nearby and trusted server. Otherwise, the communication costs could overtake the computational gain of the outsourced computations.

\end{document}